\documentclass[sigconf]{acmart}

\AtBeginDocument{%
  }

\copyrightyear{2023} 
\acmYear{2023} 
\setcopyright{rightsretained} 
\acmConference[WSDM '23]{Proceedings of the Sixteenth ACM International Conference on Web Search and Data Mining}{February 27-March 3, 2023}{Singapore, Singapore}
\acmBooktitle{Proceedings of the Sixteenth ACM International Conference on Web Search and Data Mining (WSDM '23), February 27-March 3, 2023, Singapore, Singapore}\acmDOI{10.1145/3539597.3575793}
\acmISBN{978-1-4503-9407-9/23/02}

\begin{document}

\title{Considerations for Ethical Speech Recognition Datasets}


\author{Orestis Papakyriakopoulos}
\affiliation{%
  \institution{Sony AI}
  \streetaddress{}
  \city{}
  \country{Switzerland}}
\email{orestis.papakyriakopoulos@sony.com}

\author{Alice Xiang}
\affiliation{%
  \institution{Sony AI}
  \streetaddress{}
  \city{}
  \country{USA}}
\email{alice.xiang@sony.com}

\renewcommand{\shortauthors}{Papakyriakopoulos \& Xiang}

\begin{abstract}

Speech AI Technologies are largely trained on publicly available datasets or by the massive web-crawling of speech. In both cases, data acquisition focuses on minimizing collection effort, without necessarily taking the data subjects' protection or user needs into consideration. This results to models that are not robust when used on users who deviate from the dominant demographics in the training set, discriminating individuals having different dialects, accents, speaking styles, and disfluencies.
In this talk, we use automatic speech recognition as a case study and examine the properties that ethical speech datasets should possess towards responsible AI applications. We showcase diversity issues, inclusion practices, and necessary considerations that can improve trained models, while facilitating model explainability and protecting users and  data subjects. We argue for the legal \& privacy protection of data subjects, targeted data sampling corresponding to user demographics \& needs, appropriate meta data that ensure explainability \&  accountability in cases of model failure, and the sociotechnical \& situated model design. We hope this talk can inspire researchers \& practitioners to design and use more human-centric datasets in speech technologies and other domains, in ways that empower and respect users, while improving machine learning models' robustness and utility.  
\end{abstract}



\maketitle

\section{Data \& Diversity}

The selection \& curation of datasets for model training is a fundamental step in machine learning. Not only datasets are connected with model robustness, but also to serious ethical concerns such as the privacy of individuals included, representational harms, and biases on downstream use \cite{peng2021mitigating}. Indeed, in the field of speech technologies, prior research studies have shown that commercial automated speech recognition (ASR) applications result to the discrimination of individuals that hold non-dominant dialects, accents, and speaking styles \cite{koenecke2020racial,martin2021spoken,markl2022language}, with speech recognition having a higher error rate on them. Therefore, we argue that it is important to understand the limitations of existing datasets that result to problematic model outcomes, and to develop and apply data collection \& curation practices that result to inclusive model deployment.  

Most ASR models are trained either by using open source datasets \cite{baevski2020wav2vec} or crawling the open web \cite{radford2022robust}. Nonetheless, these approaches face serious limitations when developing responsible AI applications. By reviewing the properties of over 100 open source datasets, we show that the majority of datasets contain speech that stems from dominant lingual variants, without including speech from existing dialects or accents that non-dominant social groups hold. Furthermore, datasets face serious limitations in terms of age \& gender diversity, with datasets either not including specific genders or age groups at all, or being highly unbalanced. This property results in models trained on them having poor performance on specific intersectional groups \cite{feng2021quantifying,yeung2018difficulties}. Furthermore, very few datasets contain observations of individuals having disfluent or dysarthric speech, denoting that very limited attention has been given to ASR technologies working robustly on disadvantaged social groups. Equally importantly, open source datasets contain limited diversity in terms of domain-specific speech, with content often collected from isolated and special sources (e.g. the Bible \cite{black2019cmu}, the Parliament \cite{wang2021voxpopuli}), which diverge from the sociopolitical reality of specific communities. Similarly, limited attention has been given to the diversity of recording conditions (background noise, background speech) and the recording media used (recording device properties etc.), which can have detrimental influence on the performance of ASR models and correlate with the socioeconomic status of individuals. 

The massive crawling of online-resources has been presented as a solution for many of the above issues  \cite{radford2022robust,chen2021gigaspeech}, since it is more probable to include in the used dataset a higher variation of accents, dialects, languages, and speaking styles. Nonetheless, we explain that such practice can have further negative consequences. First, either crawled speech is accompanied with captions of unknown quality, or it needs to be transcribed by the use of ASR technologies that will reproduce preexisting biases or by human annotators that might not be familiar with the accents, dialects, and speaking styles of individuals in the corpus. Second, collected data remain unbalanced and obscure, since the exact properties of the crawled speech was never available, leading to issues of accountability and explainability in cases of models' poor performance. Third, even if it is legally permissible, web data collection might violate the privacy of individuals and pose risks on them, about which they are not informed. We claim thus that there needs to be a more careful consideration about what data are used when training ASR technologies in order to mitigate unwanted harms \& consequences for users and data subjects. 
\vspace{-0.1cm}
\section{Best Practices}

Taking the limitations of existing datasets and collection practices into consideration, we describe a set of best practices that aim to mitigate harms, be inclusive to different communities and users, while highlighting the importance of interdisciplinary research and a human-centered approach when developing AI applications. These practices include following properties:

\begin{itemize}
    \item \textbf{Situated data collection.} The dataset should include observations that reflect the linguistic diversity in terms of gender, age, dialects, accents, speaking styles, recording environments and recording media that correspond to the populations that are going to be the users of the developed speech technologies. These properties should be included in the meta-data of the corpus, in order to facilitate explainability in cases of model failure. 
    \item \textbf{Sociotechnical design.} The properties of the dataset should be designed after concretizing the technical, ethical, and sociopolitical requirements of the ACR application. This means that data curators need to understand how different demographic properties of the user population will be represented in the different categories of accents, dialects, and speaking styles in the dataset. Furthermore, how the speech corpora will correspond and empower the sociopolitical background of the users in terms of vocabulary diversity and representation. 
    \item \textbf{Reflexive transcribing.} In case that the speech observations need to be transcribed, annotators should ideally match the demographics of the speakers. This will prevent misidentification of words and expressions. In case this is not possible, limitations and annotator demographics have to be reported. 
    \item \textbf{Consent, privacy \& respect of data subjects.} Data subjects should sign consent forms that explicitly describe the exact scope and context of ASR applications that are going to be trained on their data. In case datasets are to be made public, there must be explicit safeguards in the related license that will protect individuals from re-identification. Furthermore, data curators should implement appropriate mechanisms so that data-subjects information can be forgotten upon request. Dataset creators should also compensate data subjects for their work according to a country's labor policies.
    \item \textbf{Benevolent model deployment.} Datasets developed for responsible AI applications should provide explicit licensing guidelines for benevolent model usage. Dataset creators should restrict the usage of the data for biometric identification or for malicious purposes such as surveillance.
\end{itemize}

We argue that the above practices can significantly improve the performance of ASR technologies, while supporting the development of responsible AI. By centering the interests and needs of users \& data subjects, we believe that researchers and practitioners can develop datasets that are more inclusive, diverse, and empowering, while simultaneously making models more robust. We hope that our talk motivates speech and further machine learning communities to reevaluate and reflect on their practices when developing and using datasets for model training. 

\vspace{-0.1cm}

\section{Company Portrait}
As a wholly owned subsidiary of Sony Group Corporation, Sony AI was established in April 2020 to accelerate the fundamental research and development of AI and enhance human imagination and creativity, particularly in the realm of entertainment. AI Ethics sits at the core of the company's activities, developing new tools and techniques that can be used safely in critical domains such as imaging, computer vision and gaming.

\vspace{-0.2cm}

\section{Speaker Bio}
Orestis Papakyriakopoulos is a research scientist at Sony AI, where he studies the ethics of spoken and natural language processing technologies. His research showcases political issues and provides ideas, frameworks, and practical solutions towards just, inclusive and participatory socio-algorithmic ecosystems. Orestis's work has been published in conferences such as WWW, ICWSM, FAccT, AIES, EAMMO, venues in which he serves regularly as chair or program committee member. His work has been cited also by numerous news outlets such as the Financial Times and the Washington Post.

\bibliographystyle{ACM-Reference-Format}
\bibliography{bibliography}

\end{document}